\documentclass[letter,twocolumn]{jpsj3}
\usepackage{txfonts}
\usepackage{amsmath}
\usepackage{color}
\usepackage{url}

\allowdisplaybreaks

\newcommand{\BETSAu}{$\alpha$-(BETS)$_2$AuCl$_2$}
\newcommand{\ETI}{$\alpha$-(ET)$_2$I$_3$}
\newcommand{\BETSI}{$\alpha$-(BETS)$_2$I$_3$}

\title{Ambient-Pressure Organic Dirac Electron State in $\alpha$-(BETS)$_2$AuCl$_2$}

\author{Takuya Kobayashi$^{1,2}$\thanks{tkobayashi@mail.saitama-u.ac.jp}, Kazuyoshi Yoshimi$^{3}$\thanks{k-yoshimi@issp.u-tokyo.ac.jp}, Aoto Nishimoto$^{1}$, Shinji Michimura$^{1}$, and Hiromi Taniguchi$^1$}
\inst{$^1$Graduate School of Science and Engineering, Saitama University, Saitama, 338-8570, Japan \\
$^2$Research and Development Bureau, Saitama University, Saitama 338-8570, Japan \\
$^3$Institute for Solid State Physics, University of Tokyo, Kashiwa, Chiba 277-8581, Japan\\
} 

\abst{
We report an ambient-pressure Dirac electron (DE) state in a new organic conductor, $\alpha$-(BETS)$_2$AuCl$_2$ (BETS = bis(ethylenedithio)tetraselenafulvalene). 
This salt exhibits characteristic transport properties, including large positive in-plane and anomalous negative interlayer magnetoresistance. 
These signatures closely resemble the high-pressure DE states of $\alpha$-(ET)$_2$I$_3$ (ET = bis(ethylenedithio)tetrathiafulvalene). 
First-principles calculations including spin-orbit coupling identify the electronic state as a quasi-three-dimensional massive Dirac semimetal with residual Fermi pockets. This discovery provides a valuable platform for exploring bulk Dirac fermions without the complexity of high-pressure measurements.
}

\begin{document}
\maketitle

Graphene is a single-atom thick material with a two-dimensional (2D) honeycomb lattice structure, exhibiting a unique electronic property in which the energy dispersion relation becomes linear 
near the Fermi level \cite{Novoselov2005}. 
This linear dispersion gives rise to low-energy excitations that are fundamentally different from conventional parabolic-band electrons and can be described as massless Dirac fermions. 
These quasiparticles obey the 2D massless Dirac equation, representing a relativistic electronic state in which electrons and holes behave symmetrically. 
Due to the effectively zero carrier mass, graphene exhibits extremely high carrier mobility (10$^4$–10$^5$ cm$^2$/V·s), and characteristic quantum transport phenomena such as the quantum Hall effect and minimum conductivity can be observed even at room temperature \cite{Novoselov2007}. These exceptional properties have drawn considerable attention to graphene as a conductor with potential for ultrafast electronic devices.

\begin{figure}[tbp]
\begin{center}
\includegraphics[width=\columnwidth]{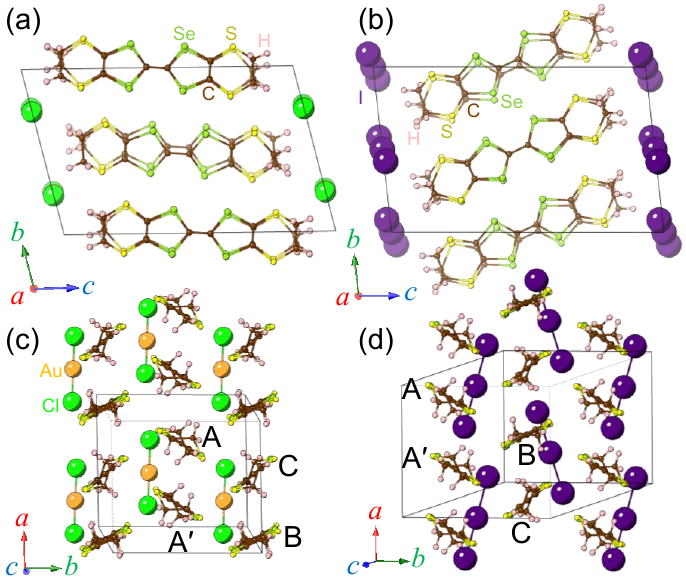}
\end{center}
\caption{
Crystal structures of (a) \BETSAu\ and (b) \BETSI\ viewed along the $a$ axis, showing alternating stacks of donor molecules and anions. Arrangement of donor molecules and anions in (c) \BETSAu\ and (d) \BETSI, viewed along the molecular long axis. Black lines indicate the unit cell. The structure of \BETSI\ was drawn using the crystallographic data deposited as CCDC 2008983 \cite{Kitou2021a}.
} 
\label{Structure}
\end{figure}

In the field of organic conductors, the compound $\alpha$-(ET)$_2$I$_3$ (ET = bis(ethylenedithio)tetrathiafulvalene) was reported early on to exhibit Dirac electron (DE) states under high pressure exceeding $1.5$~GPa \cite{Kajita1992, Tajima2000, Katayama2006, Tajima2006}, and has since been widely recognized as a pioneering example of a bulk DE system. 
More recently, an isostructural compound \BETSI\ (BETS = bis(ethylenedithio)tetraselenafulvalene) \cite{Kobayashi1993b} (Fig.~\ref{Structure}(b) and \ref{Structure}(d)), in which inner sulfur atoms in \ETI\ are replaced by selenium, has been found to host a Dirac-like state with a small band gap \cite{Kitou2021a}. 
It has been suggested that the enhanced
electron correlation
in this compound could lead to topological insulating behavior \cite{Nomoto2023}.

Thus, the realization of DE systems and topological insulating states has been demonstrated 
in organic conductors, highlighting the growing importance of exploring electronic states emerging from structural and chemical diversity. 
In this context, a key distinction between graphene and organic conductors
lies in their dimensionality. 
Although graphene is a strictly 2D material, 
organic conductors have a three-dimensional (3D) structure and exhibit 3D properties such as resistivity between layers. 
Furthermore, as bulk materials, organic DE systems enable bulk-sensitive measurements such as NMR \cite{Hiraki2011, Hirata2016,Hirata2017,Fujiyama2022}. 
It also features a tilted Dirac cone because of its low structural symmetry, which makes it particularly notable \cite{Goerbig2008,Kobayashi2009}. 
This DE state emerges upon applying pressure to a charge-ordered phase in \ETI, indicating that electronic correlations play an essential role in its realization \cite{Ohki2020, Ohki2022, Ohki2023}. 
In other words, while \BETSI\ possesses the essential characteristics of a DE system, it also exhibits complex structural and electronic diversity, resulting in unique bulk properties that distinguish it from graphene.

Importantly, \BETSI\ is the only organic conductor for which systematic physical-property measurements at ambient pressure are being pursued to test Dirac-like behavior. 
However, whether its bulk state is truly Dirac or topological-insulator-like remains unresolved.
A claim of an ambient-pressure DE state has been made for [Pt(dmdt)$_2$] \cite{Zhou2019}, but obtaining suitable single crystals is difficult and detailed investigations have not progressed.
Furthermore, the requirement of high pressure to induce the DE state in \ETI\ poses significant challenges for comprehensive measurements and potential device applications. 
Therefore, the development of organic conductors that exhibit DE states at ambient pressure is highly desirable.

Organic conductors offer high tunability through molecular design and chemical modification, making them promising platforms for creating new classes of DE systems that incorporate various additional functionalities. 
Such developments are expected to significantly expand the landscape of Dirac materials and deepen our understanding of the fundamental nature and universality of DE systems. 
Here, we introduce a new molecular charge-transfer salt, \BETSAu, and demonstrate that a quasi-3D massive Dirac semimetallic state is realized without applying pressure by transport measurements and first-principles calculations. 
We further compare its transport anisotropy with those of the $\alpha$-type DE systems and discuss the crystallographic origin of its enhanced three-dimensionality.

Single-crystal X-ray diffraction at $150$~K determined the crystal structure and lattice parameters \cite{xray} (see Supplemental Materials (SM) \cite{SM} and also Refs.~[\citen{Yamamoto2025, Sheldrick2015, Sheldrick2015a, Dolomanov2009, QE, GGA_PBE, Hamann_ONCV2013, Schlipf_CPC2015, Pizzi2020, AOYAMA2024109087}] therein, for experimental details and computational methods). 
\BETSAu\ adopts a layered structure in which BETS donor layers and AuCl$_2^-$ anion layers alternate along the $c$ axis (Fig.~\ref{Structure}(a)). 
Within a BETS layer (Fig.~\ref{Structure}(c)), columnar stacks form two sublattices: columns of A and A$^{\prime}$ molecules in general positions related to inversion symmetry, and columns formed by B and C molecules located on inversion centers. The symmetry of the donor layer is the same as in the \BETSI\ (Fig.~\ref{Structure}(d)) \cite{Kitou2021a}.

In \BETSI, the iodine atoms at the center of the I$_3^-$ anions lie on inversion centers, and its bonding direction is tilted with respect to the donor-stacking direction (Fig.~\ref{Structure}(d)). 
By contrast, in \BETSAu, the AuCl$_2^-$ anion occupies general positions, and its bonding direction is aligned nearly parallel to the donor stacks (Fig.~\ref{Structure}(c)). 
Such an arrangement implies more closely packed layers.
Actually, the (001) interplanar spacing of \BETSAu, $d_{001}=16.672$~\AA, is approximately 4.5\% shorter than that of \BETSI, $d_{001}=17.463$~\AA \cite{xrayI3}. 
Moreover, the in-plane area $ab\sin{\gamma}$ changes by only $\sim 0.2$\% whereas the unit-cell volume decreases by $\sim 4.7$\%, indicating that the volume reduction is dominated by interlayer contraction along [001]. This implies stronger three-dimensionality in \BETSAu\ than in \BETSI, without significantly altering the intralayer interactions.

\begin{figure}[tbp]
\begin{center}
\includegraphics[width=1\columnwidth]{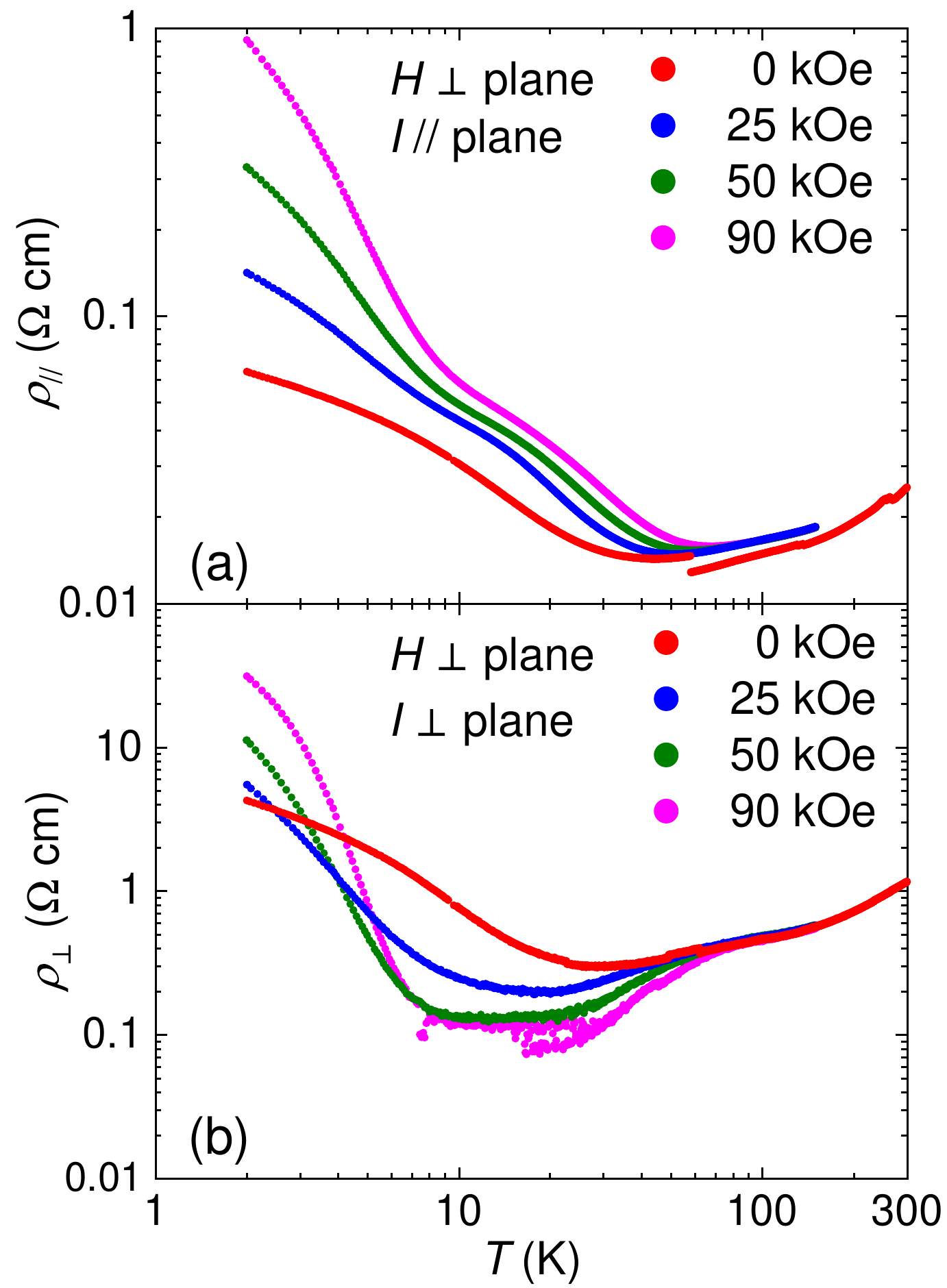}
\end{center}
\caption{
Temperature dependence of (a) in-plane resistivity $\rho_{//}$ and (b) interlayer resistivity $\rho_{\perp}$ at various magnetic fields applied perpendicular to the conducting plane. } 
\label{Resistivity}
\end{figure}

Figure~\ref{Resistivity} shows the temperature dependence of the electrical resistivity of \BETSAu\ in various magnetic fields. Panel (a) presents the in-plane resistivity, $\rho_{\parallel}$, where the current is applied parallel to the conducting plane, while panel (b) shows the interlayer resistivity, $\rho_{\perp}$, where the current is applied perpendicular to the conducting plane. The anisotropy in resistivity at room temperature, defined as $\rho_{\perp}/\rho_{\parallel}$, is  $\sim 46$, which is approximately one order of magnitude lower than that of \BETSI\ and \ETI\ at ambient pressure (both $\sim 10^3$) \cite{Sato2020, Nomoto2023}, suggesting a more 3D electronic structure, which is consistent with structural expectations.

The magnetic field is applied perpendicular to the conducting plane in all measurements. Therefore, panel (a) corresponds to the transverse magnetoresistance configuration, while panel (b) corresponds to the longitudinal magnetoresistance configuration. 
In zero magnetic field, both $\rho_{\parallel}$ and $\rho_{\perp}$ exhibit weak metallic behavior from room temperature down to $\sim 30$~K, followed by a change in temperature dependence toward weak semiconducting behavior at lower temperatures. This behavior is clearly distinct from that of \BETSI\, which shows a sharp increase in resistivity at low temperatures \cite{Nomoto2023, Inokuchi1995}, and is instead reminiscent of the behavior of $\alpha$-(ET)$_2$I$_3$ under pressure \cite{Liu2016}.

Under applied magnetic fields, \BETSAu\  exhibits characteristic behavior at low temperatures. In the transverse magnetoresistance configuration shown in Fig.~\ref{Resistivity}(a), a pronounced positive magnetoresistance appears below $\sim 60$~K. In the low-temperature, high-field regime, the magnetoresistance exceeds an order of magnitude, indicating that this system likely possesses high carrier mobility. Furthermore, under strong magnetic fields, the resistivity shows a two-step increase: a gradual rise from high temperatures down to $\sim 8$~K, followed by a sharper increase at lower temperatures. This behavior closely resembles those of \ETI\ and other DE candidates
under pressures \cite{Tajima2006, Inokuchi1993,Kobara2020}.

In contrast, the longitudinal magnetoresistance shown in Fig.~\ref{Resistivity}(b) displays an anomalous negative magnetoresistance in the intermediate temperature range, followed by a sharp increase in resistivity below $\sim 8$~K. Considering that the longitudinal configuration typically exhibits negligible magnetoresistance in conventional metals or semiconductors, such anomalous behavior cannot be explained by standard semiconductor models.

Interestingly, this temperature-dependent behavior in the longitudinal magnetoresistance is also strikingly similar to that observed in \ETI\ under pressure \cite{Tajima2009,Sugawara2010}. In that system, negative magnetoresistance at high temperatures is attributed to electron condensation into the zero-mode Landau level, while at low temperatures and high magnetic fields, the Landau level undergoes Zeeman splitting, leading to a semiconducting-like increase in resistivity. A
similar picture may provide a possible interpretation for \BETSAu. 
However, as will be discussed later in light of the band-structure calculations presented below, the rapid increase in resistivity under strong magnetic fields at low temperatures may involve a different mechanism.

In any case, the transverse and longitudinal magnetoresistance behaviors observed in \BETSAu\ differ significantly from those of conventional semiconductors and \BETSI, and instead closely resemble the characteristics of \ETI\ under pressure, where a DE system emerges. These findings strongly indicate the presence of a Dirac-like electronic state in the present compound.

\begin{figure}[tbp]
\centering

\includegraphics[width=1\columnwidth]{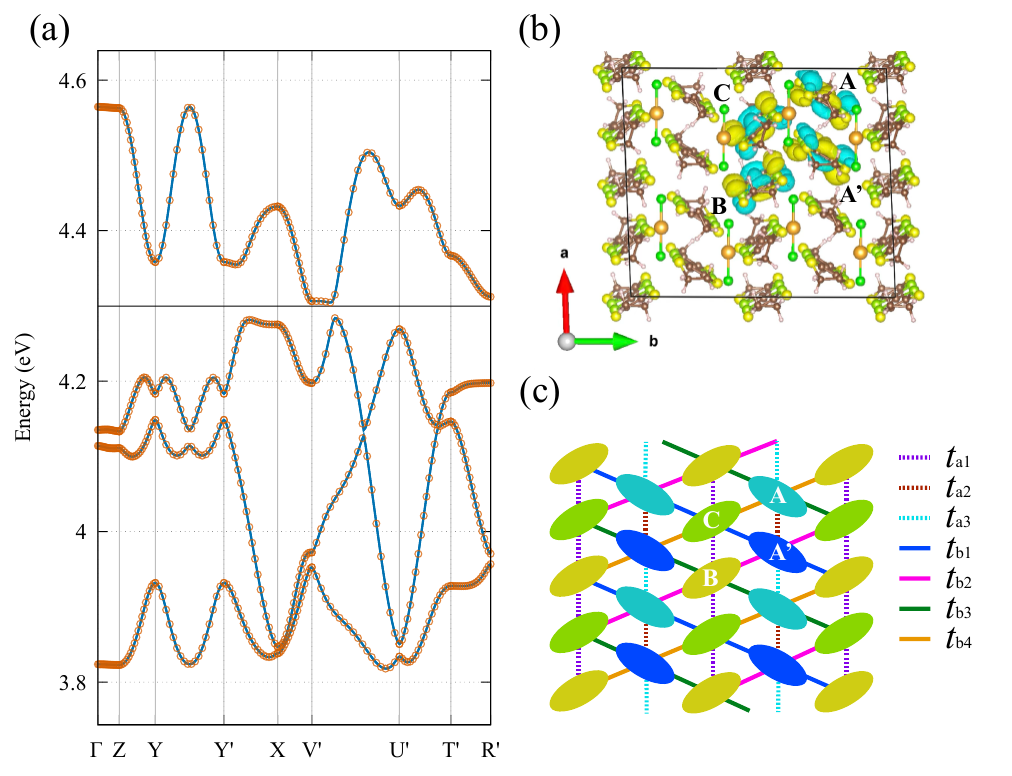}

\vspace{1em}

\begin{minipage}{\columnwidth}
\noindent
\begin{minipage}[t]{0.03\columnwidth}
(d)
\end{minipage}%
\begin{minipage}[t]{0.97\columnwidth}
\centering
{\small
\begin{tabular}{ccc}
\hline
Transfer integral & Re$(t)$ (meV) & Im$(t)$ (meV) \\
\hline
$t_{\rm a1} (\uparrow, \uparrow)$ & $9.14$  & $-0.995$\\
$t_{\rm a1} (\uparrow, \downarrow)$ & -$0.225$  & $-0.149$ \\
$t_{\rm a2}(\uparrow, \uparrow)$ & $-19.1$ & $0$ \\
$t_{\rm a2}(\uparrow, \downarrow)$ & $0$ & $0$ \\
$t_{\rm a3}(\uparrow, \uparrow)$ & $52.3$  & $0$ \\
$t_{\rm a3}(\uparrow, \downarrow)$ & $0$ & $0$ \\
\hline
$t_{\rm b1} (\uparrow, \uparrow)$ & $127$   & $- 18.0$ \\
$t_{\rm b1} (\uparrow, \downarrow)$ & $-9.03$   & $0.245$ \\
$t_{\rm b2} (\uparrow, \uparrow)$ & $150$   & $- 17.2$ \\
$t_{\rm b2} (\uparrow, \downarrow)$ & $-10.5$   & $-0.06$ \\
$t_{\rm b3} (\uparrow, \uparrow)$ & $62.5$  & $- 8.14$ \\
$t_{\rm b3} (\uparrow, \downarrow)$ & $-4.83$  & $0.847$ \\
$t_{\rm b4} (\uparrow, \uparrow)$ & $11.8$  & $- 4.14$ \\
$t_{\rm b4} (\uparrow, \downarrow)$ & $-1.82$  & $-0.67$ \\
\hline
\end{tabular}
}
\end{minipage}
\end{minipage}

\caption{
(a) Band dispersion of \BETSAu.
The symbols represent the band structure calculated by the first-principles calculations, whereas the solid curves denote the Wannier-interpolated band structure derived from MLWFs.
The Fermi energy is indicated at $E_{\rm F}=4.30$~eV. The band dispersion is plotted along the high-symmetry $k$-path defined in fractional crystal coordinates as
$\Gamma(0,0,0)$,
Z$(0,0,1/2)$,
Y$(0,1/2,0)$,
Y$^\prime(0,-1/2,0)$,
X$(1/2,0,0)$,
V$^\prime(1/2,-1/2,0)$,
U$^\prime(-1/2,0,1/2)$,
T$^\prime(0,-1/2,1/2)$,
and R$^\prime(-1/2,-1/2,1/2)$.
(b) Real-space representation of the MLWFs, visualized using the \textsc{VESTA}~\cite{MommaVESTA2011}. The black lines are guides for the eyes.
(c) Schematic model of the conducting layer.
Ellipses represent BETS molecules, and colored bonds indicate the dominant intermolecular transfer integrals ($t_{\rm a1}$--$t_{\rm a3}$ and $t_{\rm b1}$--$t_{\rm b4}$).
(d) Transfer integrals of \BETSAu\ obtained from the Wannier-based spinor tight-binding model including SOC. The arrows $\uparrow$ and $\downarrow$ denote the spin indices of the corresponding matrix elements.
}
\label{Band}
\end{figure}

To complement the experimental observations, first-principles electronic-structure calculations were performed based on the experimentally determined crystal structure of \BETSAu\ at $150$~K \cite{xray, SM}.
Figure~\ref{Band} summarizes the Wannier construction and the resulting tight-binding description: panel (a) compares the first-principles band structure with the Wannier-interpolated bands, panel (b) shows the real-space representation of the maximally localized Wannier functions (MLWFs), and panel (c) illustrates the effective tight-binding model used in the subsequent analysis.
For the stacking direction, the effective tight-binding Hamiltonian contains three intermolecular transfer integrals, while four transverse transfer integrals are present for the intermolecular paths within the conducting layer.

Although spin--orbit coupling (SOC) has often been neglected in organic conductors, its relevance was pointed out by Winter \textit{et al.}~\cite{Winter2017}.
In the present Au-containing BETS compound, the presence of a heavy element further motivates an explicit inclusion of SOC in first-principles calculations.
From the Wannier-based spinor tight-binding model, we find finite spin-mixing matrix elements in the spin-flip (up–down) sector associated with the dominant transfer integrals in the absence of SOC, as summarized in Fig.~\ref{Band}(d)~\cite{Model_parameters}
The maximum magnitude of these spin-mixing elements is $\sim 10$~meV, while the root-mean-square value is $\sim 5$~meV.
These values are comparable to the effective SOC scale previously estimated for BETS salts~\cite{Winter2017}, indicating that the present Wannier-based analysis yields a physically reasonable SOC strength.

\begin{figure}[tbp]
    \centering
    \includegraphics[width=1\linewidth]{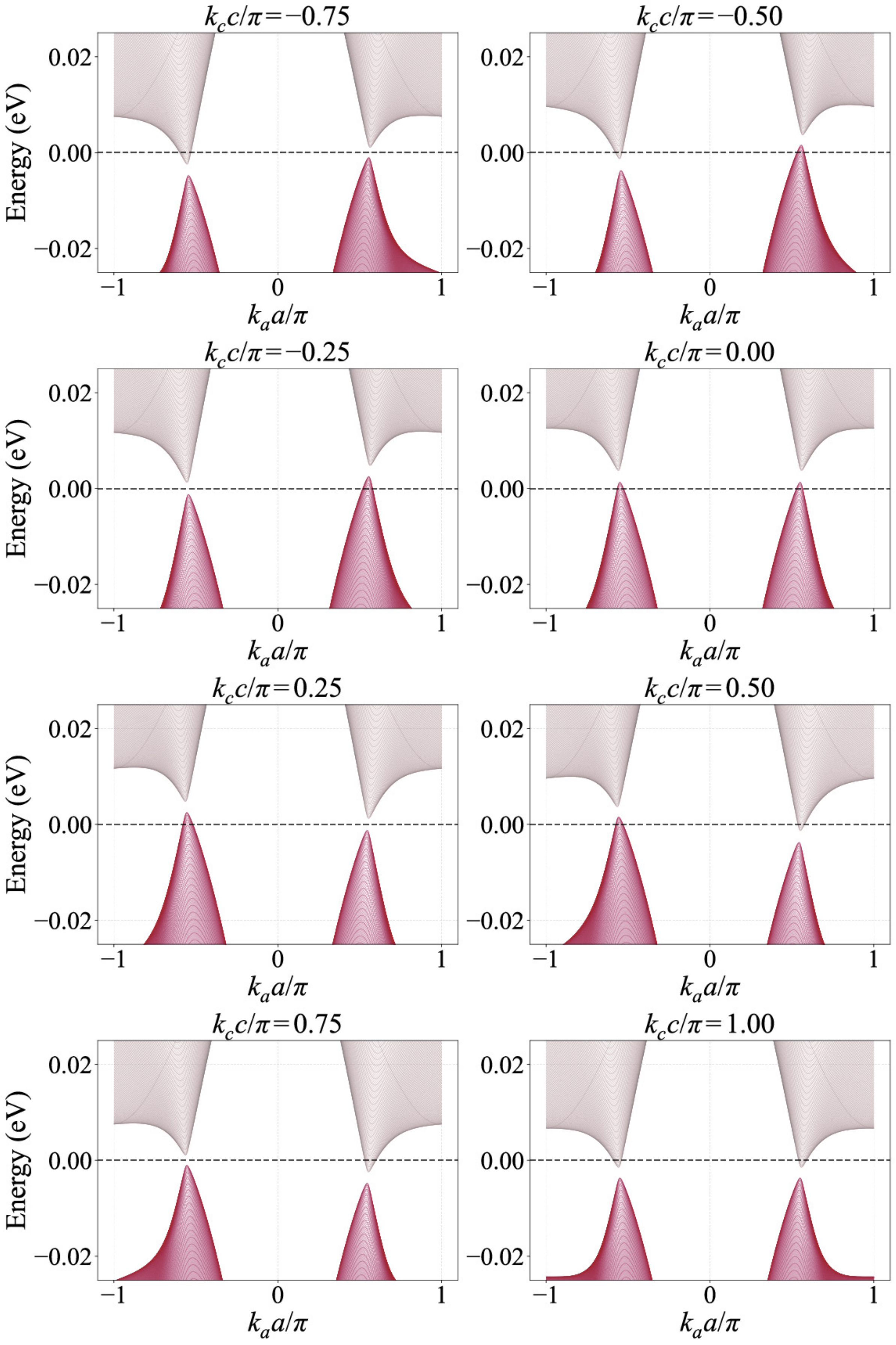}
    \caption{
Band dispersion in the presence of SOC, projected onto the $k_a$ axis for fixed values of $k_c$. Each panel shows the energy dispersion along $k_a$ at a given $k_c$, where the energy is measured relative to the Fermi energy ($E_{\rm F}=0$).}
    \label{fig:band-soi}
\end{figure}

The calculated band structures including SOC reveal the maximum transfer integral along $c$ axis,  $|t_c^{\max}|=2.28$ meV, resulting in a weak but finite dispersion perpendicular to the conducting layers.
As shown in Fig.~\ref{fig:band-soi}, SOC opens a small local gap $\Delta_{\rm SOC}$ of $\sim 3$~meV between the highest occupied band and the one immediately above in 2D slices at fixed $k_c$.
However, due to the finite interlayer dispersion, the valence and conduction bands slightly overlap along the $k_c$ direction, causing the Fermi level to intersect these bands and form small 3D Fermi pockets. 
As a consequence, \BETSAu\ realizes a \textit{quasi-3D massive Dirac semimetal}, in which a SOC-induced mass gap coexists with a residual Fermi surface originating from interlayer coupling.
For reference, the band structure without SOC is presented in Fig.~S1(a) \cite{SM}, where linear band crossings appear near the Fermi energy, highlighting the role of SOC in opening the local gap.

This electronic structure naturally accounts for the experimentally observed transport properties.
The smaller resistivity anisotropy $\rho_{\perp}/\rho_{\parallel}$ in \BETSAu\ reflects appreciable interlayer transfer integrals.
Although SOC opens a local gap, the finite interlayer dispersion bridges it, creating residual 3D Fermi pockets. 
Consequently, the system avoids a complete metal--insulator transition and instead settles into a semimetallic state at low temperatures.
By contrast, as shown in SM \cite{SM}, \BETSI\ has a more 2D electronic structure with much weaker interlayer dispersion.
In this case, the inclusion of SOC is expected to strongly suppress or even eliminate the Fermi pockets, opening a charge gap
and thereby enhancing $\rho_{\perp}/\rho_{\parallel}$.

Finally, we note that the present calculation treats the electronic structure within a single-particle framework, including SOC but neglecting explicit many-body effects.
In \BETSI, several theoretical studies have demonstrated that electron correlations, particularly on-site and intersite Coulomb interactions,
can strongly modify the Dirac spectrum and even drive interaction-induced ordered or topological phases such as the spin-ordered massive Dirac and quantum spin Hall states~\cite{Ohki2022}.
In this context, our constrained random phase approximation (cRPA) calculations using the RESPACK framework \cite{RESPACK} indicate that the magnitudes and hierarchy of the screened Coulomb interactions in \BETSAu\ are comparable to those in \BETSI\ (Table S2 in SM) \cite{SM}. This implies that electron correlation effects remain non-negligible also in \BETSAu, despite its quasi-3D electronic structure.
In \BETSAu, the residual Fermi pockets and quasi-3D dispersion may provide a qualitatively different environment for such correlation effects.
It is therefore an important future issue to clarify how the interplay between electron correlations and SOC influences the low-energy electronic states and whether correlation-driven instabilities (such as magnetic ordering, charge disproportionation, or excitonic gap formation) could emerge on top of the semimetallic background.

From the experimental perspective, systematic investigations of low-temperature magnetic and electric responses are highly valuable. 
First, further analysis of the magnetoresistance is crucial. 
In \BETSAu, negative interlayer longitudinal magnetoresistance (Fig.~\ref{Resistivity}(b)) arises even with the presence of a SOC gap, $\Delta_{\rm SOC} \sim 3~\mathrm{meV}$, and an interlayer bandwidth, $W_Z \equiv 4 |t_c^{\max}| \sim 10~\mathrm{meV}$ ($\Delta_{\rm SOC} < W_Z$). 
This will occur because the formation of the zero mode in a magnetic field increases its degeneracy, thereby enhancing the number of effective carriers contributing to interlayer conduction.
The SOC gap shifts the zero-mode energy to a finite value, and the interlayer dispersion broadens the zero mode into a finite-width one-dimensional band. However, as long as the temperature and bandwidth allow for occupation of the zero mode, the effect of magnetic-field-induced degeneracy enhancement dominates. As a result, the interlayer resistivity decreases with the magnetic field.

On the other hand, the origin of the rapid increase in resistivity under strong magnetic fields at low temperatures
may differ depending on the system. Unlike \ETI, the present material is a gapped weakly 3D system.  
The Zeeman effect at low temperatures can combine with the SOC gap and interlayer dispersion to produce an effective gap in the zero mode. Therefore, the origin of positive magnetoresistance at low temperatures in \ETI\ under pressures \cite{Tajima2009,Sugawara2010} and \BETSAu\ may differ. Further detailed studies of the magnetoresistance, including both magnetic-field and temperature dependence as well as angular dependence, are required.

Furthermore, precise quantum-oscillation and thermoelectric measurements would help to map the residual Fermi surfaces and identify many-body renormalization effects. NMR\cite{Hiraki2011,Fujiyama2022}  and $\mu$SR measurements can probe possible spin or orbital ordering, while infrared and Raman spectroscopy can track correlation-induced renormalization of the band structure. Most of these techniques take full advantage of the fact that the material exhibits DE states at ambient pressure.

In summary, we have demonstrated that \BETSAu\ realizes a DE state at ambient pressure. 
Structural analysis reveals that the alignment of AuCl$_2$ anions enhances the interlayer coupling, resulting in a significantly reduced resistivity anisotropy compared to \BETSI. 
The transport properties exhibit characteristic signatures of a DE system, specifically the large positive transverse magnetoresistance and the anomalous negative longitudinal magnetoresistance, which are remarkably similar to those of the high-pressure phase of $\alpha$-(ET)$_2$I$_3$. 
First-principles calculations including SOC support these observations, identifying the electronic state as a quasi-3D massive Dirac semimetal wherein a SOC-induced gap coexists with residual Fermi pockets. 
The discovery of this ambient-pressure Dirac material opens new avenues for exploring the interplay between relativistic electrons and strong correlations without the constraints of high-pressure environments.

\section*{Acknowledgement}
We thank N. Tajima at Toho University for valuable discussion and helpful comments. 
Single-crystal X-ray diffraction analysis was performed using a Bruker D8 QUEST ECO diffractometer installed at the Comprehensive Analysis Center for Science, Saitama University. 
The computation in this work has been done using the facilities of the Supercomputer Center, the Institute for Solid State Physics, the University of Tokyo (ISSPkyodo-SC-2024-Cb-0046, 2025-Ca-0011).
This work was partially supported by the Japan Society for the Promotion of Science KAKENHI Grant Numbers 24K17002 and 25H01403.

T.K. and K.Y. contributed equally to this work.

\section*{Data availability statement}
Crystallographic data for \BETSAu\ can be obtained free of charge from the joint Cambridge Crystallographic Data Centre (CCDC) and the Fachinformationszentrum Karlsruhe Access Structures service (\url{https://www.ccdc.cam.ac.uk/structures/}) 
under the deposition No. 2433895. 
The data supporting the findings of this article are openly available at \url{http://isspns-gitlab.issp.u-tokyo.ac.jp/k-yoshimi/alpha-bets2-aucl2}.

\bibliographystyle{jpsj}

\end{document}


\maketitle
\vspace{-0.4cm}

\section*{Experimental Details}
Single crystals of \BETSAu\ were prepared by electrochemical oxidation of BETS in chlorobenzene containing $5$\% methanol, in the presence of the supporting electrolyte tetrabutylammonium-AuCl$_2$. 
A recent, independently reported study synthesized an isostructural compound using benzonitrile as the solvent \cite{Yamamoto2025}.
Single-crystal x-ray diffraction measurements were carried out at $150$~K using a Bruker D8 QUEST ECO diffractometer with Mo-K$\alpha$ radiation ($\lambda = 0.71073$~\AA). 
The structure was solved by direct methods and refined by full-matrix least-squares using SHELXT \cite{Sheldrick2015a} and SHELXL \cite{Sheldrick2015}, respectively, as implemented in the Olex2 program \cite{Dolomanov2009}.
Electrical resistivity was measured with a standard four-probe method in a Quantum Design Physical Property Measurement System (PPMS). 
Gold wires (25 $\mu$m in diameter) were attached with carbon paste to the crystals with the dimensions of  $455 \times 850 \times 6$~$\mu$m$^3$ (for out-of-plane resistivity measurement) and $415 \times 220 \times 19$~$\mu$m$^3$ (for interlayer resistivity).

\section*{Computational Methods}
All calculations were carried out using the \textsc{Quantum~ESPRESSO} package~\cite{QE} within the generalized gradient approximation of Perdew, Burke, and Ernzerhof~\cite{GGA_PBE}, employing optimized norm-conserving Vanderbilt pseudopotentials for all atoms~\cite{Hamann_ONCV2013, Schlipf_CPC2015}.
Fully relativistic pseudopotentials were used to include spin–orbit coupling (SOC).
The kinetic-energy cutoffs for the plane-wave basis and charge density were set to 80~Ry and 320~Ry, respectively, and a $5\times5\times3$ Monkhorst--Pack $k$-point mesh was used to sample the Brillouin zone.
Hydrogen positions were relaxed without SOC, while all other coordinates were fixed to the experimental structure. All subsequent calculations were performed on this relaxed geometry, with and without SOC.

Bloch states near the Fermi level were projected onto four maximally localized Wannier functions (MLWFs) centered on each BETS molecule using the \textsc{Wannier90} code~\cite{Pizzi2020}, yielding an effective tight-binding
Hamiltonian that includes both interlayer transfer integrals and SOC terms.
The resulting tight-binding Hamiltonian was further analyzed using the \textsc{H-wave} package~\cite{AOYAMA2024109087}, where the Fermi energy was determined by fixing the total electron number of the tight-binding model.

\section*{Model Parameters in the Absence of Spin–Orbit Coupling}
\begin{figure}
  \centering
  \begin{minipage}[t]{0.48\linewidth}
    \centering
    \includegraphics[width=\linewidth]{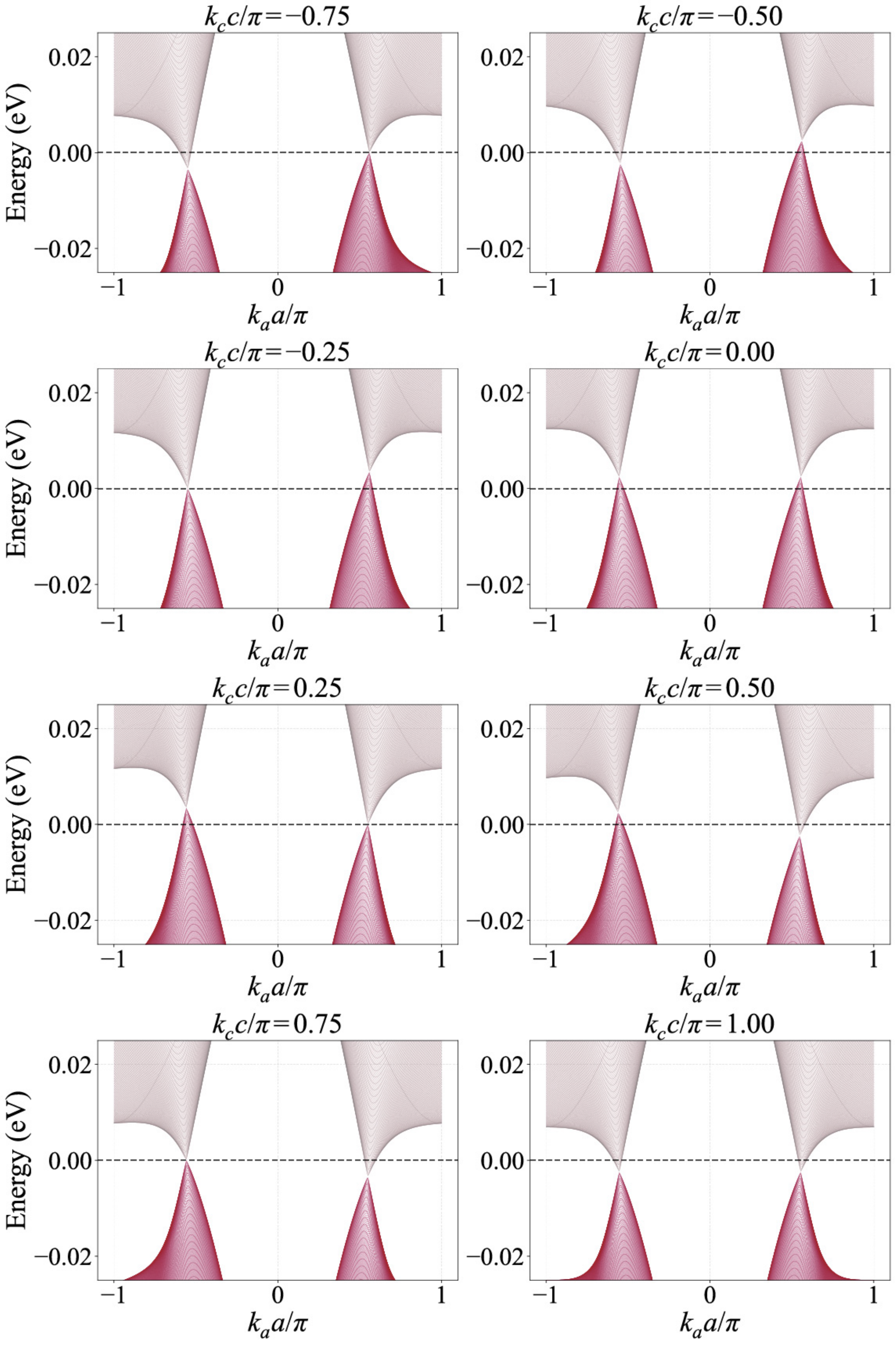}
    \small (a) AuCl$_2$
  \end{minipage}\hfill
  \begin{minipage}[t]{0.48\linewidth}
    \centering
    \includegraphics[width=\linewidth]{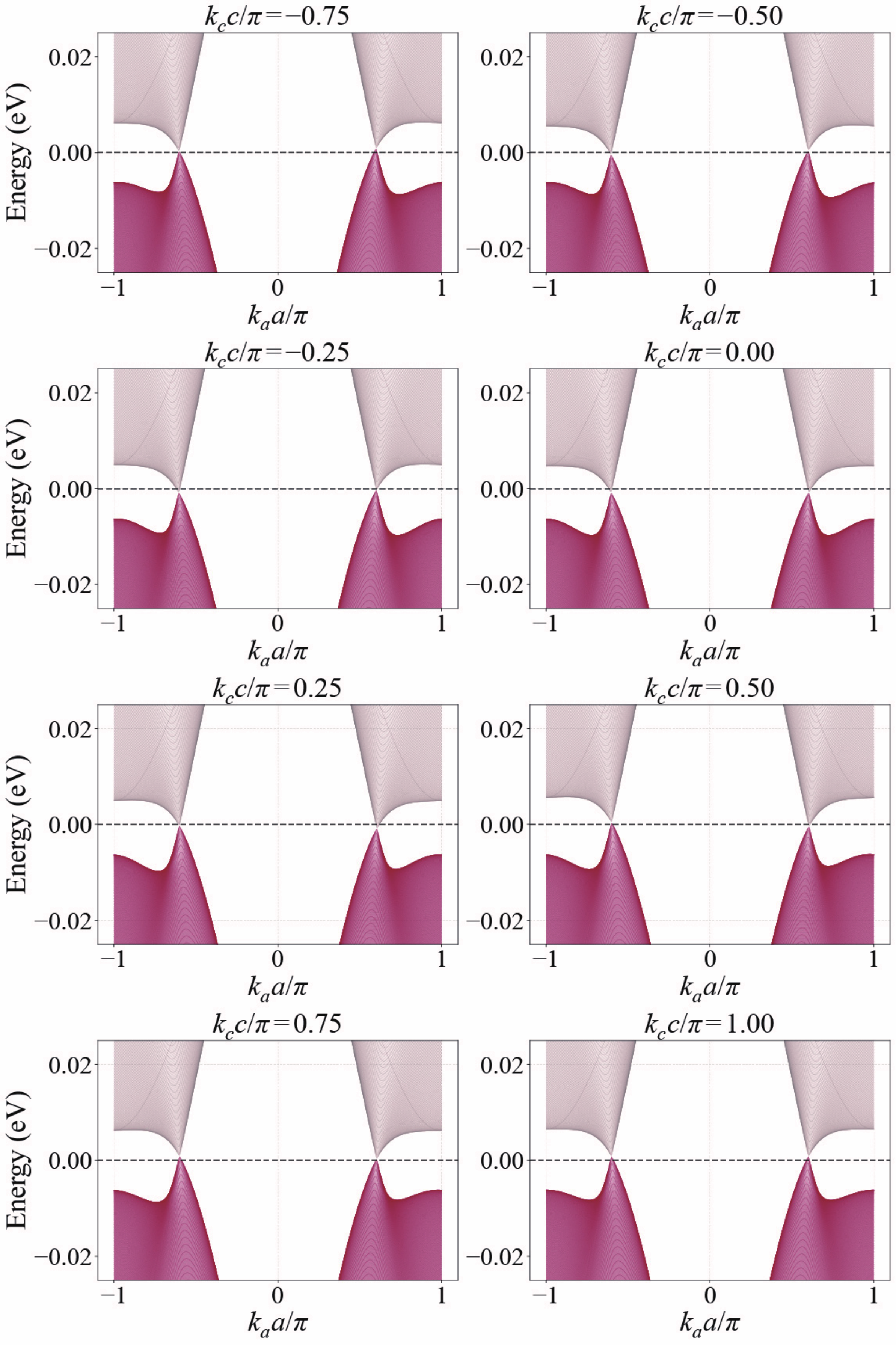}
    \small (b) I$_3$
  \end{minipage}
\caption{
Band dispersion in the absence of SOC for (a) \BETSAu\ and (b) \BETSI, projected onto the $k_a$ axis for fixed values of $k_c$.
In each panel, the energy dispersion along $k_a$ at a given $k_c$ is shown, where the energy is measured relative to the Fermi energy ($E_{\rm F}=0$).
}
  \label{fig:band-non-soi-comparison}
\end{figure}

To clarify the role of SOC in \BETSAu\
we evaluated the electronic parameters in the absence of SOC. 
The corresponding transfer integrals were extracted from the effective Hamiltonian generated using the RESPACK code~\cite{RESPACK} as shown in Table~\ref {tbl:ti-nosoc}.
We also evaluated screened Coulomb interactions within the constrained random phase approximation (cRPA) using the RESPACK code~\cite{RESPACK} as shown in Table~\ref {tbl:crpa-nosoc}. 
For \BETSI, the transfer integrals and screened Coulomb interaction parameters listed in Tables~\ref{tbl:ti-nosoc} and \ref{tbl:crpa-nosoc} were taken from Ref.~\citen{Ohki2023}, where the structure at $30$~K was used \cite{Kitou2021a}. The model parameters and the band dispersion shown in Fig.~\ref{fig:band-non-soi-comparison}(b) for \BETSI\ are available in the data repository \cite{alphasalts_repo}.

\begin{table}[h]
\centering
\begin{tabular}{crr}
\hline
Parameters & AuCl$_2$ & I$_3$\\
\hline
$t_{\rm a1}$ & 9.42  & 9.90 \\
$t_{\rm a2}$ & -18.9 & -16.4 \\
$t_{\rm a3}$ & 52.6  & 51.1 \\
$t_{\rm b1}$ & 129   & 138 \\
$t_{\rm b2}$ & 152   & 158 \\
$t_{\rm b3}$ & 63.2  & 65.7 \\
$t_{\rm b4}$ & 12.3  & 18.6 \\
\hline
\end{tabular}
\caption{Transfer integrals (in meV) obtained in the absence of SOC. 
The parameters represent the matrix elements of the effective tight-binding 
Hamiltonian constructed from the MLWFs generated using RESPACK. 
Values for \BETSI\ are taken from Ref.~\citen{Ohki2023}.}
\label{tbl:ti-nosoc}
\end{table}

\begin{table}[ht]
\centering
\begin{tabular}{crr}
\hline
Parameters & AuCl$_2$ & I$_3$\\
\hline
$U_{\rm A}$ & 1.41 &1.39\\
$U_{\rm A'}$& 1.41 &1.39\\
$U_{\rm B}$ & 1.41 &1.41\\
$U_{\rm C}$ & 1.38 &1.36\\
$V_{\rm a1}$ &  0.595 & 0.581\\
$V_{\rm a2}$ &  0.610 & 0.596\\
$V_{\rm a3}$ &  0.582 & 0.567\\
$V_{\rm b1}$ &  0.583 & 0.580\\
$V_{\rm b2}$ &  0.580 & 0.573\\
$V_{\rm b3}$ &  0.549 & 0.538\\
$V_{\rm b4}$ &  0.566 & 0.557\\
\hline
\end{tabular}
\caption{Screened onsite and intersite Coulomb interactions (in eV) obtained by cRPA in the absence of SOC using RESPACK. Values for \BETSI\ are taken from Ref.~\citen{Ohki2023}.}
\label{tbl:crpa-nosoc}
\end{table}

The transfer integrals in Table~\ref{tbl:ti-nosoc} correspond to matrix elements of the effective tight-binding Hamiltonian between symmetry-inequivalent MLWFs. 
$t_{\rm a1}$--$t_{\rm a3}$ denote transfer integrals along the stacking  direction, whereas $t_{\rm b1}$--$t_{\rm b4}$ represent transverse intermolecular paths. 
These parameters constitute the effective 2D molecular Hubbard model commonly used as a minimal description for $\alpha$-type salts.
To complement the analysis based on the effective Hamiltonian parameters, 
Fig.~\ref{fig:band-non-soi-comparison} shows the band dispersions calculated in the absence of SOC for (a) \BETSAu\ and (b) \BETSI, 
projected onto the $k_a$ axis for several fixed values of $k_c$. 
In both salts, the low-energy electronic structure near the Fermi level is characterized by linear band crossings appearing at generic $k_a$ points, 
and the overall dispersion shapes are very similar across different $k_c$ slices. 
This similarity indicates that the basic band topology and quasi-two-dimensional electronic structure are essentially preserved between the two compounds when SOC is neglected.
A closer inspection, however, reveals subtle but systematic differences between the two salts. 
Compared with \BETSAu, the band dispersion in \BETSI\ exhibits a slightly reduced bandwidth and a marginally weaker $k_c$ dependence. 

\begin{figure}[t]
  \centering
  \begin{minipage}{0.49\linewidth}
    \centering
    \includegraphics[width=\linewidth]{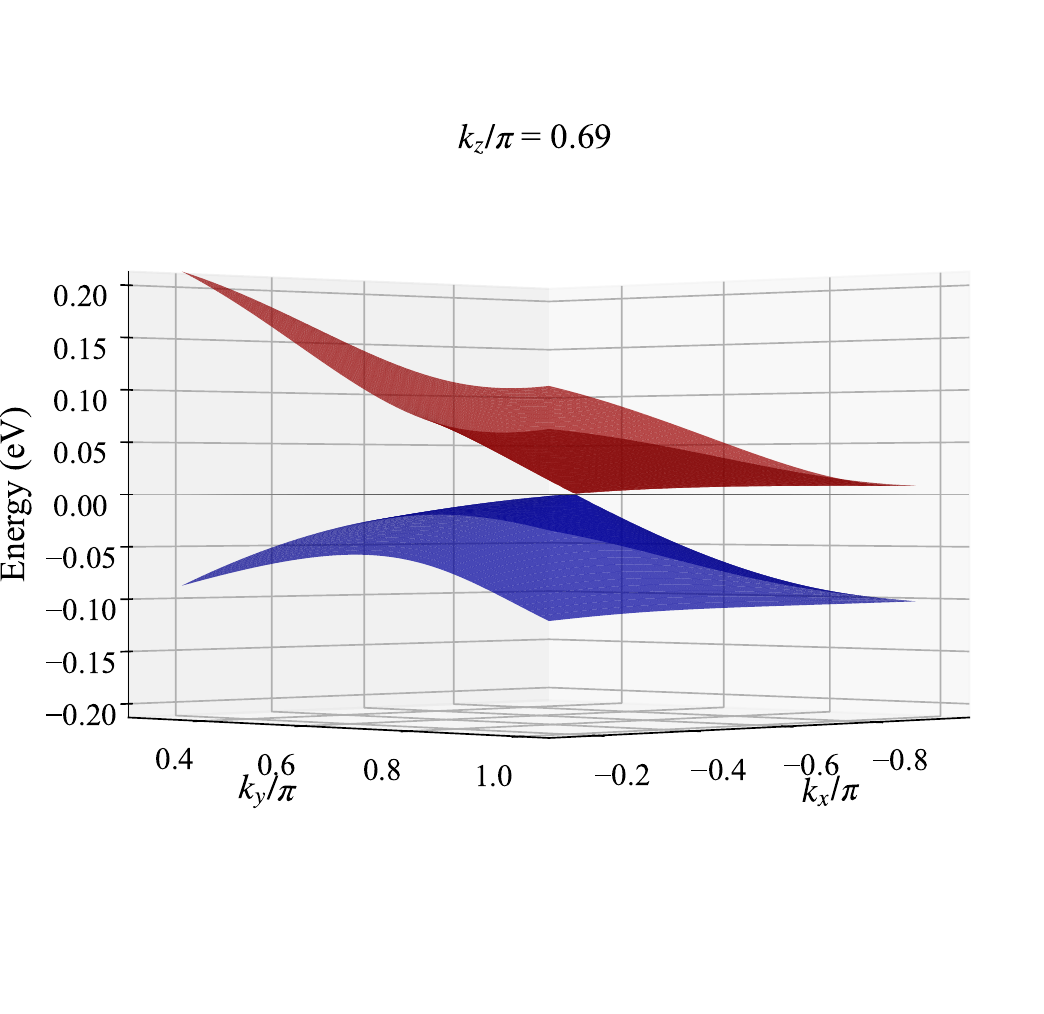}
  \end{minipage}\hfill
  \begin{minipage}{0.49\linewidth}
    \centering
    \includegraphics[width=\linewidth]{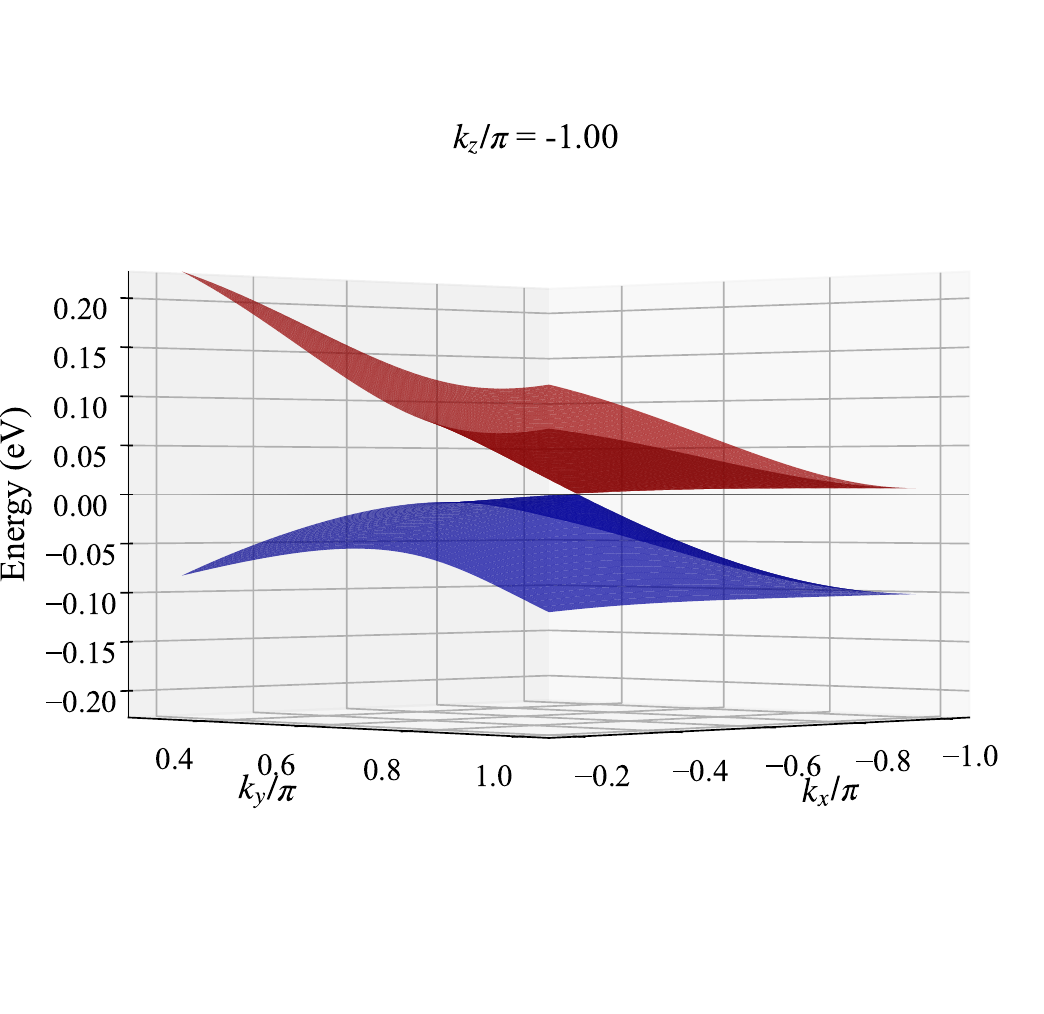}
  \end{minipage}
  \caption{
Band dispersions in the $k_x$--$k_y$ plane around the two symmetry-related Dirac points at fixed $k_z$.
Panels (a) and (b) correspond to \BETSAu\ and \BETSI, respectively.
The energy is measured relative to the Fermi level $E_\mathrm{F}$.}
  \label{fig:dirac_kxky_two}
\end{figure}

We identified the Dirac points by locating momentum points where the energy separation between the two relevant bands is minimized, and then selecting the points with the smallest energy difference from the Fermi level as approximate Dirac points.
The underlying band energies were computed on a $256 \times 256 \times 32$ $k$-mesh and subsequently interpolated onto a finer grid of $401 \times 401$ points in the $k_x$--$k_y$ plane and 201 points along the $k_z$ direction, on which the search was performed.
For \BETSAu, the Dirac points are located around $\mathbf{k}_{\mathrm{D},\pm} = \pm(-0.56, 0.70, 0.69)\pi$.
For comparison, we also analyzed the Dirac points in the iodine compound \BETSI, where two symmetry-related Dirac points were identified at $\mathbf{k}_{\mathrm{D},\pm} = \pm(-0.60, 0.70, -1.00)\pi$.
The linear band dispersions around these Dirac points in the $k_x$--$k_y$ plane at fixed $k_z$ are shown in Fig.~\ref{fig:dirac_kxky_two}.

The Fermi velocity $v_\mathrm{F}$ was evaluated from the band dispersion near the Dirac points using numerical differentiation,
\begin{equation}
  v_{\mathrm{F},\alpha} = \frac{1}{\hbar}
  \left| \frac{\partial E}{\partial k_\alpha} \right|_{\mathbf{k}=\mathbf{k}_\mathrm{D}},
\end{equation}
where $\alpha = x, y, z$.
For \BETSAu, we obtained $v_{\mathrm{F},x} \sim 3.0 \times 10^4$~m/s, $v_{\mathrm{F},y} \sim 8.4 \times 10^4$~m/s, and $v_{\mathrm{F},z} \sim 5.4 \times 10^3$~m/s.
By contrast, for \BETSI, we obtained $v_{\mathrm{F},x} \sim 2.9 \times 10^4$~m/s, $v_{\mathrm{F},y} \sim 9.2 \times 10^4$~m/s, and $v_{\mathrm{F},z} \sim 5.9 \times 10^2$~m/s.
While the in-plane Fermi velocities are comparable between the two compounds, the velocity along the $k_z$ direction is strongly suppressed in \BETSI, indicating a much weaker interlayer dispersion.

Table~\ref{tbl:crpa-nosoc} summarizes the onsite and intersite screened Coulomb interactions obtained by cRPA. $U_{\rm A}$, $U_{\rm A'}$, $U_{\rm B}$, and $U_{\rm C}$ denote the onsite interactions, whereas $V_{\rm a1}$--$V_{\rm b4}$ represent the corresponding intersite interactions defined on the same intermolecular bonds as the transfer integrals. 
Overall, the \BETSI\ exhibits slightly smaller onsite and intersite interactions compared to the \BETSAu, while preserving the same hierarchy of interaction strengths. 


\bibliographystyle{jpsj}